\documentclass[]{aa} 
\usepackage{multicol}
\usepackage{graphicx}
\usepackage{subfigure}
\graphicspath{ {images/} }

\usepackage{txfonts}
\usepackage{booktabs}
\usepackage{amsmath}
\usepackage{epstopdf}
\usepackage{enumitem}
\usepackage{multicol}
\usepackage{multirow}
\usepackage{chngcntr}

\usepackage{natbib}
\begin{document}

\title{The radio properties of the OH megamaser galaxy IRAS 02524+2046}

   \subtitle{}

   \author{ Hao Peng
          \inst{1}
          \and
          Zhongzu Wu\inst{1}\fnmsep\thanks{zzwu08@gmail.com}
           \and
         Bo Zhang\inst{2}
         \and
         Yongjun Chen\inst{2}
          \and
         Xingwu Zheng\inst{3}
          \and
         Dongrong Jiang\inst{2}
          \and
         Zhiqiang Shen\inst{2}
          \and
         Xi Chen\inst{4}
         \and
         Sotnikova, Yu. V.\inst{5}
          }

   \institute{College of Physics, Guizhou University, 550025 Guiyang, PR China
              \email{zzwu08@gmail.com}
              \and
              Shanghai Astronomical Observatory,
Chinese Academy of Sciences, 80 Nandan Road, Shanghai 200030, PR China \and
School of Astronomy and Space Science, Nanjing University, 22 Hankou Road, Nanjing 210093, PR China \and
Center for Astrophysics, GuangZhou University, Guangzhou 510006, PR China \and
Special Astrophysical Observatory, Russian Academy of Sciences, 369167, Nizhnii Arkhyz, Russia
         }

   \date{  }


  \abstract
   {
   We present results from VLBI observations of continuum and OH line emission in IRAS 02524+2046 and also arcsecond-scale radio properties of this galaxy using VLA archive data.  We found that there is no significant detection of radio continuum emission from VLBI observations. The arcsecond-scale radio images of this source show no clear extended emission, the total radio flux density at L and C band are around 2.9 mJy and 1.0 mJy respectively, which indicate a steep radio spectral index between the two band. Steep spectral index, low brightness temperature and high $q$-ratio (the FIR to the radio flux density), which are three critical indicators in classification of radio activity in the nuclei of galaxies, are all consistent with the classification of this source as a starburst galaxy from its optical spectrum. The high-resolution line profile show that both of the 1665 and 1667 MHz OH maser line have been detected which show three and two clear components respectively. The channel maps show that the maser emission are distributed in a region $\sim$ 210 pc $\times$ 90 pc, the detected maser components at different region show similar double spectral feature, which might be an evidence that this galaxy is at a stage of major merger as seen from the optical morphology.
}


  \authorrunning{Peng Hao et al. }            
   \titlerunning{RADIO EMISSION FROM IRAS 02524+2046}  
   \maketitle
%

\section{Introduction}
OH Megamasers (OHMs) are rare, luminous masers found in (ultra-)luminous infrared galaxies ([U]LIRGs) which represent different phases in the evolution of gas-rich mergers \citep{2019MNRAS.486.3350S,2020AAS...23520730R}. IRAS 02524+2046 (hearafter IRAS 02524) is a luminous infrared galaxy (LIRG) and one of
the most luminous known OH megamaser galaxies \citep{2013ApJ...774...35M,2002AJ....124..100D,2006AJ....132.2596D} with a redshift z = 0.1814. This object have the most unusual spectrum of the Arecibo OHM survey, showing multiple strong narrow components in both OH 1667 MHz (OH1667) line and OH 1665 MHz (OH1665) line.   The optical morphology of this source is elliptical-like with a single tidal tail \citep{2005MNRAS.364...99V}, which indicate that this galaxy might be also at one stage of merging process, and the optical spectrum is typical of a starburst galaxy \citep{2005MNRAS.364...99V}.

   The results from Arecibo telescope observations of IRAS 02524 show
   strong variability in many spectral components in the OH lines, suggesting that variable features are smaller than 1 pc
(0.3 mas) with a brightness temperature $T_{b}$ > $8 \times 10^{11}$ K  \citep{2007IAUS..242..417D}.
   Generally there are two types of OH megamaser emission, one is the diffuse low-gain maser emission with the emitting population of OH inverted by photons from the infrared continuum, another is the compact high-gain emission observed by VLBI, arising from the inverted population which could be collisionally pumped \citep{2005ASPC..340..224P}.
High-resolution radio continuum studies play a key role in directly imaging and determining the nature of the nuclear power sources (AGNs and/or starbursts), and the OHM studies can enhance our understanding of the kinematics on parsec scales in the nuclear region \citep{2006ApJ...653.1172M}. So far, we found that the radio properties of this source is barely available in the literature.

We observed this source with VLBA for both the OH line and continuum emission on February 17, 2017. Our main aim is to study the radio properties of this source and its high resolution OH line emission distribution.  The details about our observations and archive radio data collection are presented in Section 2.
     The results and discussions are presented in Section 3 and 4 respectively.
     In Section 5 we present the conclusions.
   Throughout the paper, cosmological parameter values of
     $H_0$ = 73 km s$^{-1}$ Mpc$^{-1}$ , and $\Omega_{matter}$ = 0.27, and $\Omega_{vacuum}$ =0.73 are adopted.

\section{Observations and data reduction}
\label{sect:Obs}

\subsection{VLBI observations}

     We performed continuum and OH line observations with VLBA on 2017 February 17 for 8 hours, including all ten VLBA antennas.
     We used J0259+1925 (located $\sim$ 1$^{\circ}$ from IRAS 02524) as the phase referencing calibrator.
    We also analysed archival data with VLBA (BD0075, P.I.: Darling J.) and EVN (GD018, P.I.: Diamond P.) experiments, the two projects also made both the continuum and OH line observations.
     The VLBA project BD0075 was performed on 2001 October 14,
     the observation spanned 12 hours with 8 hours integration time on IRAS 02524, including all ten VLBA antennas.
     The EVN project GD018 was performed on 2005 March 4, including 11 antennas, Jodrell Bank(JB), Westerbork(WB), Effelsberg(EF), Onsala(ON), Medicina (MC), Noto(NT), Torun(TR), Urumqi(UR), VLA(Y), Green Bank(GB) and Arecibo(AR). The basic information about the three VLBI observations  are listed in Table. \ref{vlbilist}.


\begin{table*}
       \caption{Parameters of the high resolution observations. }
     \label{vlbilist}
  \centering
  \begin{tabular}{c c c c l c c}     
  \hline\hline
    Observing Date & Frequencies & Array          & Phase       & Program  \\
                  &    (GHz)    & Configuration  & Calibrator  &          \\
      \hline
     2001Oct14  &  1.4 & VLBA  & J0257+1847 & DA0075 \\
     2005Mar04  &  1.4 & EVN   & J0257+1847 & GD018  \\
     2017Feb17  &  1.4 & VLBA  & J0259+1925 & BC229A\\
      \hline
       \end{tabular}
   \end{table*}


\begin{table*}
       \caption{Parameters of the VLA. }
     \label{vladata}
  \centering
  \begin{tabular}{c c c c l c c}     
  \hline\hline
    Observing Date & Frequencies & Array          & Program & Integrated Flux & Map peak \\
                  &    (GHz)    & Configuration    &         &   (mJy)         &(mJy/beam) \\
      \hline
     1993Nov01  &  1.4 & VLA-D     & NVSS    & 2.600$\pm$0.500   &  2.21   \\
     2002Jan25  &  4.9 & VLA-A    & AD0461  & 1.077$\pm$0.136   & 0.58\\
     2003Aug04  &  1.4 & VLA-A        & AD0483  & 2.990$\pm$0.463   & 2.74\\
     2005Mar04  &  1.4 & VLA-B          & GD0018  & 3.104$\pm$0.332   & 3.20\\

      \hline
       \end{tabular}
   \end{table*}


\subsection{VLA archival data}
In order to study the radio properties of this source, we also used the VLA archive data. There are four VLA
    Observations available for IRAS02524 and the list of VLA observations is summerised in Table. \ref{vladata}.  The project name 'NVSS' in this table stands for NRAO VLA Sky Survey\footnote{https://www.cv.nrao.edu/nvss/}.
\subsection{Data reduction}
  The VLBA, EVN and VLA data in this paper are all calibrated using the NRAO Astronomical Image Processing System (AIPS) package. The initial calibration of VLBA and EVN data follows a online Tutorial\footnote{http://www.aoc.nrao.edu/~amiodusz/sumschool14/vlbaspectut.shtml}, the main procedures include ionospheric correction, amplitude calibration, editing, instrumental phase corrections, antenna-based fringe-fitting of the phase calibrator and then apply the solutions to the target source, for the velocities of the OH lines we have also corrected the effects of the Earth's rotation and its motion within the Solar System.  The calibration of VLA data follows standard procedures in AIPS. We have flagged the channels that may contain OH emission lines for making the images of the continuum emission.  After the calibration in AIPS, we imported the calibrated data into the DIFMAP package \citep{1995BAAS...27..903S} to obtain the continuum and OH line images. Because the weakness of continuum and line emission gives rise to a low signal- to-noise ratio (S/N), no self-calibration was performed.
\begin{figure}
   \centering
\includegraphics[width=8cm]{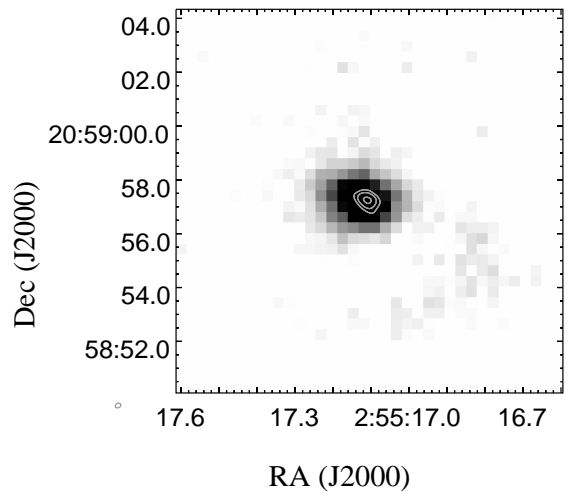}
      \caption{The VLA C-band radio contour image overlaid on SDSS R-band grey image of IRAS 02524. The pixel scale of SDSS image is 0.396''/pixel, the center of this map is : RA: 02 55 17.104, Dec: +20 58 57.210. The image parameters of the radio continuum image is as follows: peak 0.000579 Jy/beam, beam width: 468 $\times$ 433 (mas) at $2.27^{\circ}$  and the contours is 0.000135 Jy/beam $\times$ (1 2 4).}
      
    \label{vlac}%
    \end{figure}

\section{Results}

\subsection{Radio continuum emission in IRAS 02524}

    We have reduced our VLBA project and other two VLBI observations listed in Table. \ref{vlbilist}, no significant continuum emission was detected at a 3 $\sigma$ level. In order to study the continuum emission, we have only included the OH emission line-free channels, the sensitivities of the three observations for continuum emission range from around 10 $\mu$Jy/beam to 50 $\mu$Jy/beam.

    The VLA observations of this source are listed in Table. \ref{vladata}, there are three epoch observations at L band and one epoch at C band.  The image from VLA-A array at C band overlaid on the SDSS R-
band grey image of this source is presented in Fig. \ref{vlac} which shows that the radio continuum emission has similar emitting regions as in optical, other images are presented in Fig. \ref{vlal}, these images all show no significant extend structure. The peak and Integrated flux density are present in Table. \ref{vladata}, the L band flux density of this source is around 2.9 mJy, and C band around 1.1 mJy. The spectral indexes $\alpha_{total}$ and $\alpha_{peak}$ derived by using the total flux and peak flux, are around -0.8 and -1.2 respectively.

    We have also observed this source with RATAN-600 radio telescope on Dec. 2019, our preliminary results show that the radio flux was not detected at all observed frequencies.  The most sensitive RATAN radiometer is at 4.7 GHz, we estimated the flux density upper limit at this frequency is equal to 1 mJy, which is consistent with our VLA C-band result in Table. \ref{vladata}. The RATAN-600 observations are present in more detail in Section \ref{ratanobs}.

\begin{figure}
   \centering
   \includegraphics[width=8cm]{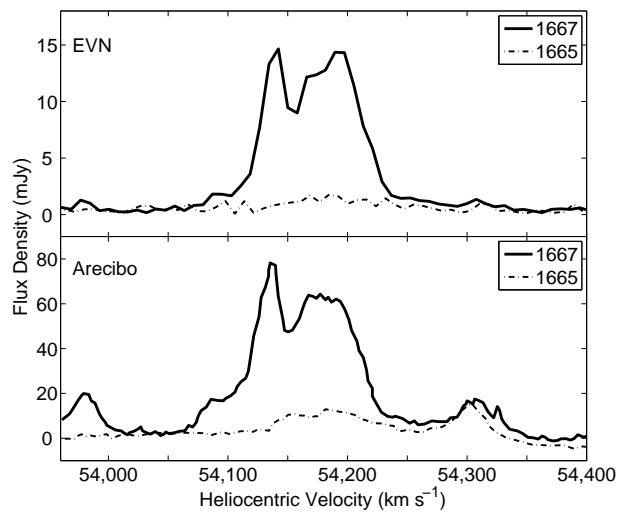}
      \caption{The vector averaged cross-power OH maser spectra of IRAS 02524. The upper panel is the high \textbf{angular} resolution VLBI spectra inspected from EVN project GD018; The bottom panel is the spectra from \cite{2013ApJ...763....8M}, which is derived from single-dish observations by Arecibo telescope.
      }
         \label{evnspectrum}
   \end{figure}

\subsection{High \textbf{angular} resolution OH line emission}
Our VLBA project (BC229B) show no clear spectrum from the calibrated visibility data, then we investigated the velocity profile of the spectrum from the calibrated EVN project GD018, which are present in Fig. \ref{evnspectrum}. We can see that  both of the OH main lines(1667 MHz line and 1665 MHz line) are detected. The profiles are consistent with the single dish line profiles observed by Arecibo telescope \citep{2013ApJ...763....8M}, with $\sim$ 20\% flux density of the single-dish observations restored.

  \begin{figure*}
   \centering
   \includegraphics[width=18cm]{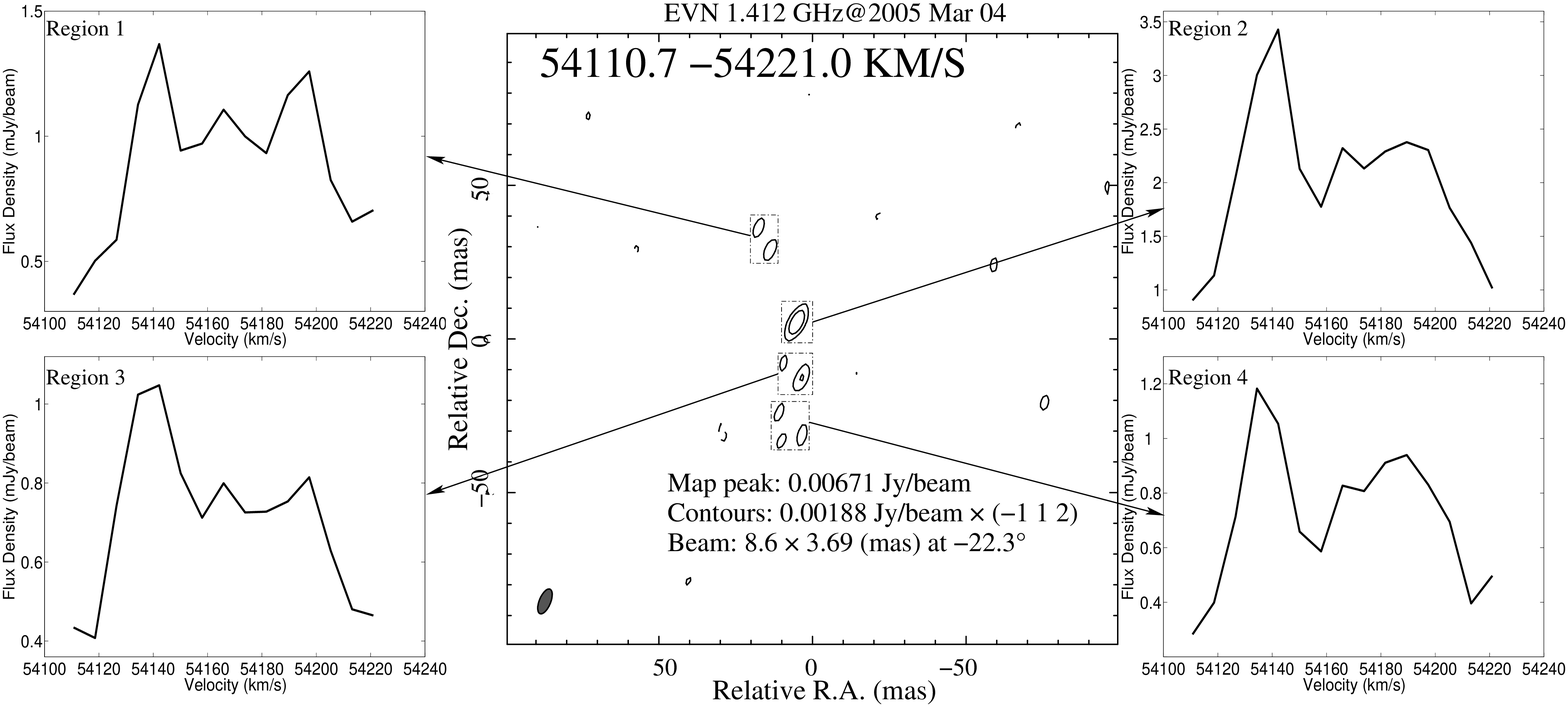}
      \caption{Combined channel map of the OH 1667 MHz line emission in IRAS 02524 observed with the EVN project GD018.
      }
     \label{1667combin}%
    \end{figure*}

    \begin{figure*}
   \centering

   \includegraphics[width=14cm]{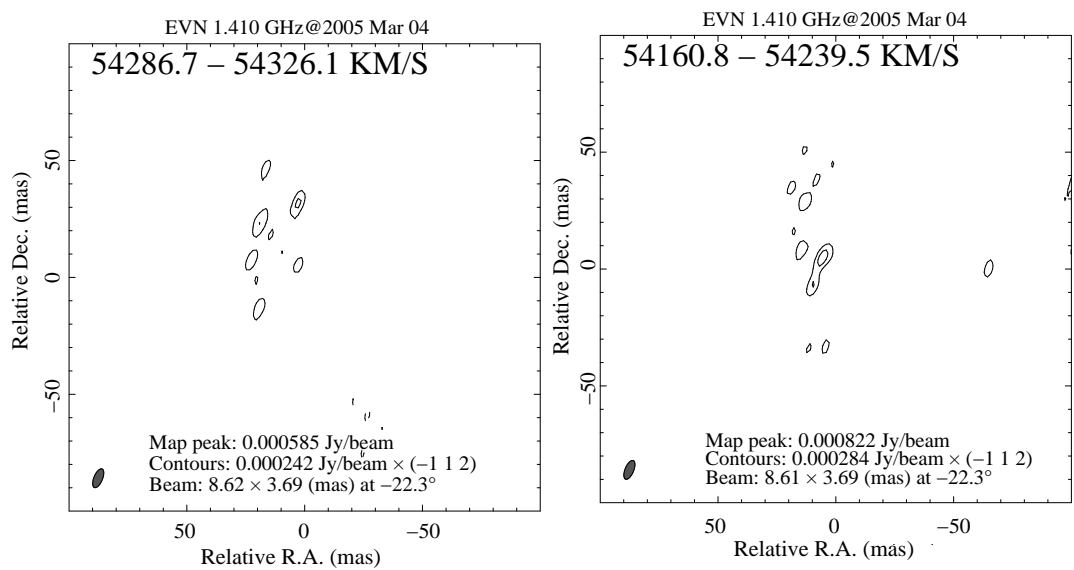}
      \caption{Combined channel map of the OH 1665 MHz line emission in IRAS 02524 observed with the EVN project GD018.
      }
     \label{1665combin}%
    \end{figure*}

Fig. \ref{evnspectrum} shows three evident components in the 1667 MHz lines: 54291.9 - 54323.4 km s$^{-1}$, 54110.7 - 54221.0 km s$^{-1}$, and 53961.0 - 54000.4 km s$^{-1}$. We have made the images of the three components by combining the channels together, the images are present in Fig. \ref{1667combin} and Fig. \ref{1667three}. We also made channel maps for each channel from 54110.7 - 54221.0 km s$^{-1}$, these maps show similar structure to the combined channel map, which are presented in Fig. \ref{channelmap1}. The flux density of OH1665 line is much lower than the OH1667 line, we combined the channels from 54160.8 - 54239.5 km s$^{-1}$ and 54286.7 - 54326.1 km s$^{-1}$, and the combined channel maps are present in Fig. \ref{1665combin}. We can see that the maser emission are resolved into a number of components and distributed in similar region around 70 mas (210 pc) in the north-east direction and less than 30 mas (90 pc) in the east-west direction.

 The two VLBA observations listed in Table. \ref{vlbilist} have much lower baseline sensitivities than the EVN project GD018(joint observation of EVN and large radio telescopes, see Section 2.1),  both of the projects show no evident spectrum. We averaged channels in the frequency range from 1411.5 MHz to 1412.5 MHz, which correspond to the 1667 MHz line component from 54110.7 - 54221.0 km s$^{-1}$, the images are presented in Fig. \ref{2vlbaline}, we found that only the brightest component in Fig.  \ref{1667combin} is detected.

\section{Discussions}{\label{sec:discussion}}
\subsection{The radio continuum}
We found that there are no significant continuum emission above a 3 $\sigma$ level with the VLBI observations listed in Table. \ref{vlbilist}. While the dirty map of EVN project GD018(see Fig. \ref{dirtymap}), showed that there are tentatively some possible continuum emission at the center of the map with a peak around 50$\sim$60 uJy/beam, while it also looks like noise due
to data reduction, we estimated that the flux density upper limit of VLBI observations of this source is around 50$\sim$60 uJy/beam which can yield a brightness temperature to be $\sim$ 1 $\times$ $10^{6}$K by using the beam size of this observation, this is consistent with a starburst origin similar to that seen in IRAS 12032+1707 \citep{2005ApJ...618..705P}. Further radio observations of sufficiently high sensitivity are needed to study the high-resolution radio continuum structure of this source. These results indicate that no compact emission of high-brightness temperature are present in this galaxy . We found  that the $\alpha_{total}$ and $\alpha_{peak}$ are around -0.8 and -1.2 respectively, which consistent with optically thin synchrotron radiation and characteristic of normal spirals, extended starbursts and compact starbursts in ULIRGs\citep{2015ApJ...799...25S}. The brightness temperature ($T_B$) of the continuum emission at C-band with VLA observation (project AD0461) is around $\sim 5 \times 10^{2}$ K.  These results agree with the classification of this source as starburst galaxy from its optical spectrum.


There are also other parameters used in the literature to classify AGN and starburst galaxies, one is the logarithm of the ratio of the FIR to the radio flux density ($q$-ratio) \citep{2001ApJ...554..803Y}, which suggest that with decrease of q, contribution from AGN to the total radio emission becomes more significant. Specifically,
   the $q$-ratio is defined as $q = log [3.36 \times 10^{2} (2.58 S _{60 \mu m} + S _{100 \mu m})/S_{radio}]$ \citep{1985ApJ...298L...7H,1991ApJ...378...65C},
   where $S _{60 \mu m}$, $S _{100 \mu m}$ and $S_{radio}$ are IRAS 60 $\mu$ m, 100 $\mu$ m flux and radio flux density respectively.
 IRAS 02524 has been detected by IRAS at 60 $\mu$m with a flux density of 0.958 Jy,
   and at 100 $\mu$m with a flux density of $<$ 4.79 Jy \citep{2002AJ....124..100D}. Plus the VLA fluxes in Table. \ref{vladata}, the $q$ value for
   IRAS 02524 is estimated to be 2.92 at 1.415 GHz,
   and $q\simeq$ 3.26 at 4.86 GHz, which is much larger than the threshold of $q$ < 2.50 \citep{2006A&A...449..559B} to be an AGN.

 $T_B$, $\alpha$ and $q$-ratio are three critical parameters in classification of radio activity in the nuclei of galaxies,  \cite{2006A&A...449..559B} defined an activity factor : $\beta_{c} = 0.308(q - 0.75 \ast T_{B} + 3 \ast \alpha - 1)$, which represents a weighted qualification of the three diagnostic parameters with different weights.
 Starburst-like source have a positive value of $\beta_{c}$, and AGN-like sources have negative values. We calculate the value of $\beta_{c}$ for IRAS 02524, which is around 1.32, indicating that it is much more like a starburst-like source.

   Using a starburst-dominated sample, \cite{2017MNRAS.471.1634H} showed that the Infrared luminosity $L_{IR}$ (between 8 and 1000 $\mu$m) and radio luminosity at 1.4 GHz are both the star-formation rate(SFR) indicators, and the relations are as follows: SFR = 4.5 $\times$ $10^{-44}$ $L_{IR}$ and SFR =1.02 $\times$$10^{-28}$$L_{1.4 GHz}$. IRAS 02524 has been detected by IRAS at 60 $\mu$m with a flux
density of 0.958 Jy \citep{2002AJ....124..100D}, then we calculated the SFR from $L_{IR}$ to be 199.7 $M_{\odot}$ $yr^{-1}$. Such a SFR would produce a non-thermal radio flux 2.15 mJy at 1.4 GHz. The L band radio flux of this source from VLA observations is around 2.9 mJy (see Table. \ref{vladata}), which favors that the radio continuum emission in this galaxy is probably dominated by starburst.
\subsection{The OHM emission}
In this paper, we have detected both the OH1667 line and OH1665 line emission with VLBI, it seems that all the line components found in single-dish observations are available in our VLBI OH line spectrum(see Fig. \ref{evnspectrum}), but only with around 20 \% of the single dish flux density restored. This means that all the components are probably with some contributions from compact maser clouds. Based on Arecibo telescope observations, \cite{2002ApJ...572..810D} found that the hyperfine ratios (1667-1665 MHz flux density ratio) for individual narrow features in the spectrum are $R_{H}$ = 1.4, 5.63, 1.88 from high velocity to low, while \cite{2013ApJ...774...35M} found that there are only two peaks in the OH1665 line emission that align with peaks in the OH1667 line emission. We found that VLBI OH line spectrum also only show two peaks in the OH1665 line emission which is consistent with the results in \cite{2013ApJ...774...35M}. We calculated the $R_{H}$ for the two components are around 1.81 and 8.99 from high to low velocity (see Fig. \ref{evnspectrum}).

The VLBI images of maser line emission (see Fig. \ref{1667combin}-\ref{1665combin} and Fig. \ref{1667three}-\ref{channelmap1})  show that the components are all distributed in an area less than 70 mas (210 pc) in the north-east direction and less than 30 mas (90 pc) in the east-west direction.  \cite{2007IAUS..242..417D} found that IRAS 02524 show strong variability in many spectral components in the OH lines, suggesting that variable features are smaller than 1 pc (0.3 mas) with a brightness temperature $T_{b}$ > $8 \times 10^{11}$ K. We tried to fit the brightest components in the channel maps, we found that the size of the brightest component is around 3 mas in size, and $T_{b}$ is $\sim$ $4 \times 10^{9}$ K, which indicate that the maser components might be still compact and needs higher angular resolution observations.

The OHM emission profile is supposed to be attributed to an ensemble of many masing regions, \cite{2002ApJ...569...87L} suggested that the variability of OH line emission might be caused by the compact maser emission through interstellar medium, yielding an interstellar scintillation. The three epoch VLBI images (Fig. \ref{1667combin} and \ref{2vlbaline}) of this source all show one dominant bright component. We can see that the peak flux density of three observations are different, this might be caused by the variability of this component. Moreover, we also can see that the EVN image (Fig. \ref{1667combin}) also show some weak emission regions.  We think that strong variability found by \cite{2007IAUS..242..417D} might arise from the contributions from the brightest component showed in the VLBI images and some other analogous compact maser components which
are not detectable due to the limited sensitivity, because the VLBI scale flux density seems only one-fifth of the single-dish Arecibo observation (see Fig. \ref{evnspectrum}).  

Fig. \ref{1667combin} show that all the four regions of the OH1667 brightest line component display a similar double spectral feature, suggesting the existence of a velocity substructure in all the regions. \cite{2016ApJ...825..128L} have presented a classification scheme of merging stage of LIRGs, single nuclei and one obvious tidal of this source \citep{2005MNRAS.364...99V} indicate that this source is possibly at a stage of major merger. Because the largest size among the regions are around 210 pc, the existence of a velocity substructure in all the regions support the view that this galaxy might be at a merging stage and the gas clouds might originate from two or more galaxies.

\subsection{The nature of the radio emission in OHM galaxies}
In general, OHM line emission have been observed in (U)LIRGs and it is likely that OH megamasers occur during a specific state, or stage of the merger, which are consequences of tidal density enhancements accompanying galaxy interactions \citep{2007IAUS..242..417D,Pihlstrom2007}. The hosts of OHM galaxies usually present features of starburst and AGN, and the radio continuum emission are produced from these phenomena. There are two possible explanations for these features:

 \begin{itemize}

 \item The OHM galaxies could represent a transition stage between a starburst and the emergence of an AGN \citep{2006AJ....132.2596D}.
 One possible scheme might be that in the initial merging stage of OHM formation, the background radio continuum mainly comes from the starburst and FIR dominates the population inversion, the radio continuum and OHM structure are supposed to be relatively diffuse; as galaxy merging proceeds, AGN activity in producing radio continuum
and inverting population ground level becomes more and more significant, the radio spectrum gets
relatively flat, the structure of radio and OHMs emission becomes relatively compact, and the
OHMs velocity fields will become more ordered just as shown in Mrk 231 \citep{Klockner2003}. Our results show that IRAS 02524 do not show clearly circumnuclear structure or velocity gradients, and the radio continuum emission is consistent with starburst origin, which mean that this source might at the early stage of merging process under this simplified view of OHMs.

 \item
The OHM galaxies might originate in a central AGN, contaminated by emission of circumnuclear star-forming regions \citep{2018MNRAS.474.5319H,2018MNRAS.479.3966H}. Although the radio continuum emission of many OHMs seems to show a link with nuclear starburst, while a hidden AGN might also be possible as the LIRGs that host OH megamasers tend to be of infrared excess compared with those that do not \citep{2003MNRAS.342..373R,2002AJ....124..100D}. Our results show that the radio emission of IRAS 02524 also seems to be produced by nuclear starburst, and it seems no evident contributions from the AGN, while also we can not exclude the existence of a hidden AGN, further high sensitivity and high spatial resolution optical spectrum observations of the nuclear region might be useful to find the existence of an hidden AGN in this galaxy \citep{2018MNRAS.474.5319H}.

 \end{itemize}
Because the OHM galaxies are related with merging processes, it may contain two or more intervening systems or regions with much different properties. In order to understand the nature of radio emission in OHM galaxies and further test about the above scenarios, high sensitivity interferometry and VLBI observations of a large number of OHM galaxies will help to determine whether they are hosting an AGN or compact starburst, their connections with the merging stages and other environment parameters. Meanwhile, the high resolution optical-IR multi-wavelength imaging and integral field spectroscopy analysis of large sample of OHM galaxies are also important for understanding the nature of the nuclear region and then the origin of radio emission in OHM galaxies \citep{2019MNRAS.486.3350S,2018MNRAS.474.5319H,2018MNRAS.479.3966H}. 


\section{Summary}
We have presented the radio properties of IRAS 02524 from VLBA, EVN and VLA observations. We found that there is no significant continuum emission detectioned above 3 $\sigma$ level from three epoch VLBI observations. The L and C band integrated radio flux from VLA observations of this source is around 2.9 mJy and 1.1 mJy and the spectral indexes, $\alpha_{total}$ and $\alpha_{peak}$ are around -0.8 and -1.2 respectively. The brightness temperature of this source is around 500 K from VLA C-band observation.  The L and C band $\rm q$-ratio is  $\sim$2.92
and  $\sim$3.26, respectivley, which is much larger than the threshold of $q$ < 2.50 typical of an AGN.   These parameters all show that this source is probably a starburst galaxy, which is consistent with the classification from optical spectrum of this source.

The high-resolution radio spectrum of this source show both the OH1667 line and OH1665 line, the profile is similar to the single dish spectrum. The channel maps show that the emission of the OH maser lines located in a region around 210 pc $\times$ 90 pc region. We also showed that the detected maser clouds at different region have similar double spectral feature, which suggesting a velocity substructure in all the detected regions, which might be an evidence that this galaxy is in the process of merging as seen from the optical morphology.
%


\begin{acknowledgements}
We thank the anonymous referee for his/her helpful suggestions and comments.
This work is supported by the grants of NSFC(
Grant No. 11763002, U1931203) and the Major Program of NSFC(Grant No.
11590780, 11590784). The European VLBI Network
is a joint facility of European, Chinese, and other radio astronomy
institutes funded by their national research councils. The National
Radio Astronomy Observatory is operated by Associated Universities,
Inc., under cooperative agreement with the National Science
Foundation.
\end{acknowledgements}

\bibliography{penghao}


\appendix
\counterwithin{figure}{section}
\section{Online materials}
\subsection{RATAN observations}
\label{ratanobs}
RATAN observations were carried out in December 2019 in the transit mode. Details of observations and data analysis can be found in \cite{2016AstBu..71..496U,2017AN....338..700M}. The measurements were made at six frequencies simultaneously: 1.28, 2.25, 4.7, 8.2, 11.2 and 21.7 GHz. Parameters of the continuum radiometers are given in the Table. \ref{tab:radiometers}. The detection limit for RATAN single sector is approximately 5 mJy/beam under good conditions at 4.7 GHz and at an average antenna elevation. At other frequencies this value depends on the atmospheric extinction and the effective area on the certain antenna elevation.

      The number of IRAS 02524+2046 observations was 14 in December 2019 and the radio flux was not detected at the most sensitive RATAN radiometer at 4.7 GHz. The flux density upper limit at this frequency is equal to 0.001 Jy. Due to a strong Radio Frequency Interference (RFI) the bandwidths at 1.28 and 2.25 GHz are quite narrow (80 and 60 MHz, respectively) and the fluxes density at these frequencies have not detected also. Only one measurement in the radio continuum is currently known for IRAS 02524+2046 at 1.4 GHz and it is equal to 2.6 mJy \citep{1998AJ....115.1693C}. Probably it is possible to get the radio signal at 4.7 GHz with the RATAN using more long-term accumulation.

\begin{table*}
\caption{\label{tab:radiometers}RATAN-600 continuum radiometers.
Where $f$$_0$ -- central frequency, $\Delta$$f_0$ -- bandwidth,
$\Delta$$F$ -- flux density detection limit per beam,
and BW -- beam width -- an angular resolution in RA (an angular resolution in declination is three to five times worse than in RA).}
\centering
\begin{tabular}{rlcr}
\hline
 $f_{0}$ & $\Delta$$f_{0}$ & $\Delta$$F$ &  BW \\
  GHz    &   GHz           &  mJy/beam   &  arcsec \\
\hline
$21.7$	& $2.5$	& $50$	&	11	\\
$11.2$	& $1.4$	& $15$	&	15.5	\\
$8.2$		& $1.0$	& $10$	&	22	\\
$4.7$		& $0.6$	& $5$		&	35	\\
$2.25$	& $0.08$	& $40$	&	80	\\
$1.28$	& $0.06$	& $200$	&	110	\\
\hline
\end{tabular}
\end{table*}
\begin{figure*}[htbp]
   \centering
   \includegraphics[width=18cm]{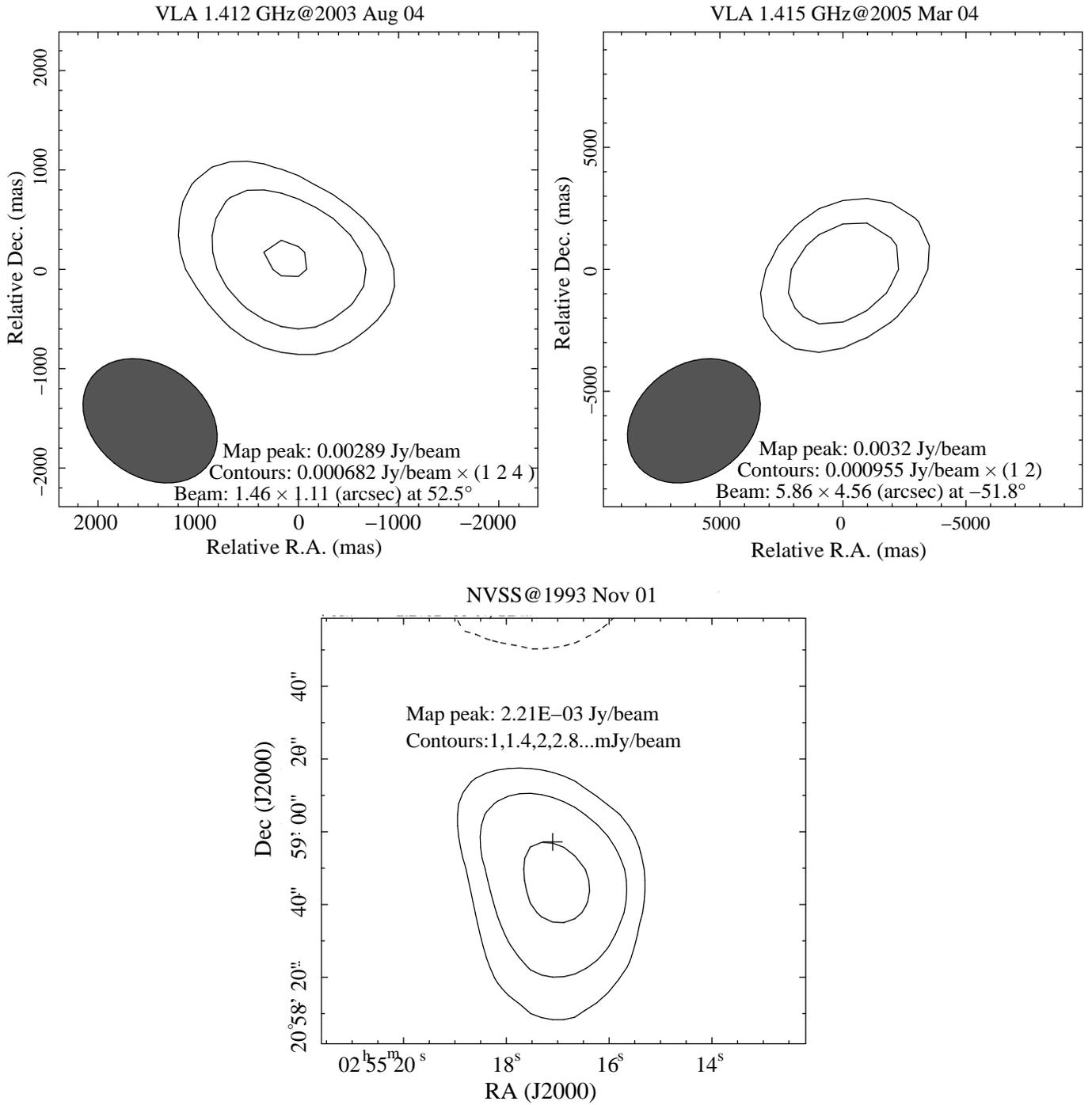}
      \caption{The VLA L-band radio contour images of IRAS 02524.
      }
    \label{vlal}%
    \end{figure*}

  \begin{figure*}
   \centering
   \includegraphics[width=18cm]{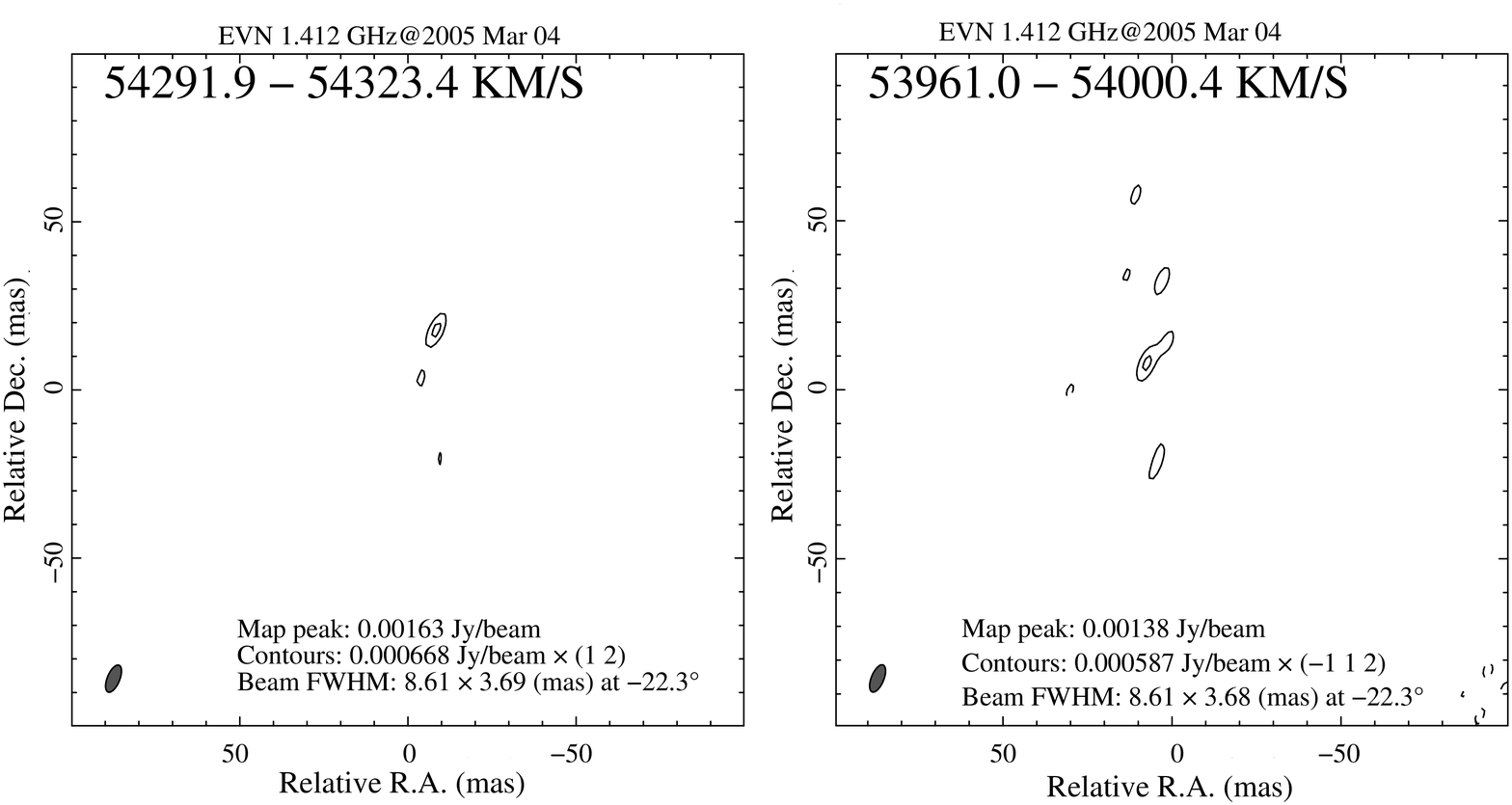}
      \caption{Combined channel maps of the OH 1667 MHz line emission in IRAS 02524 observed with the EVN project GD018.
      }
     \label{1667three}%
    \end{figure*}

   \begin{figure*}[htbp]
   \centering
   \includegraphics[width=17cm,height=24cm]{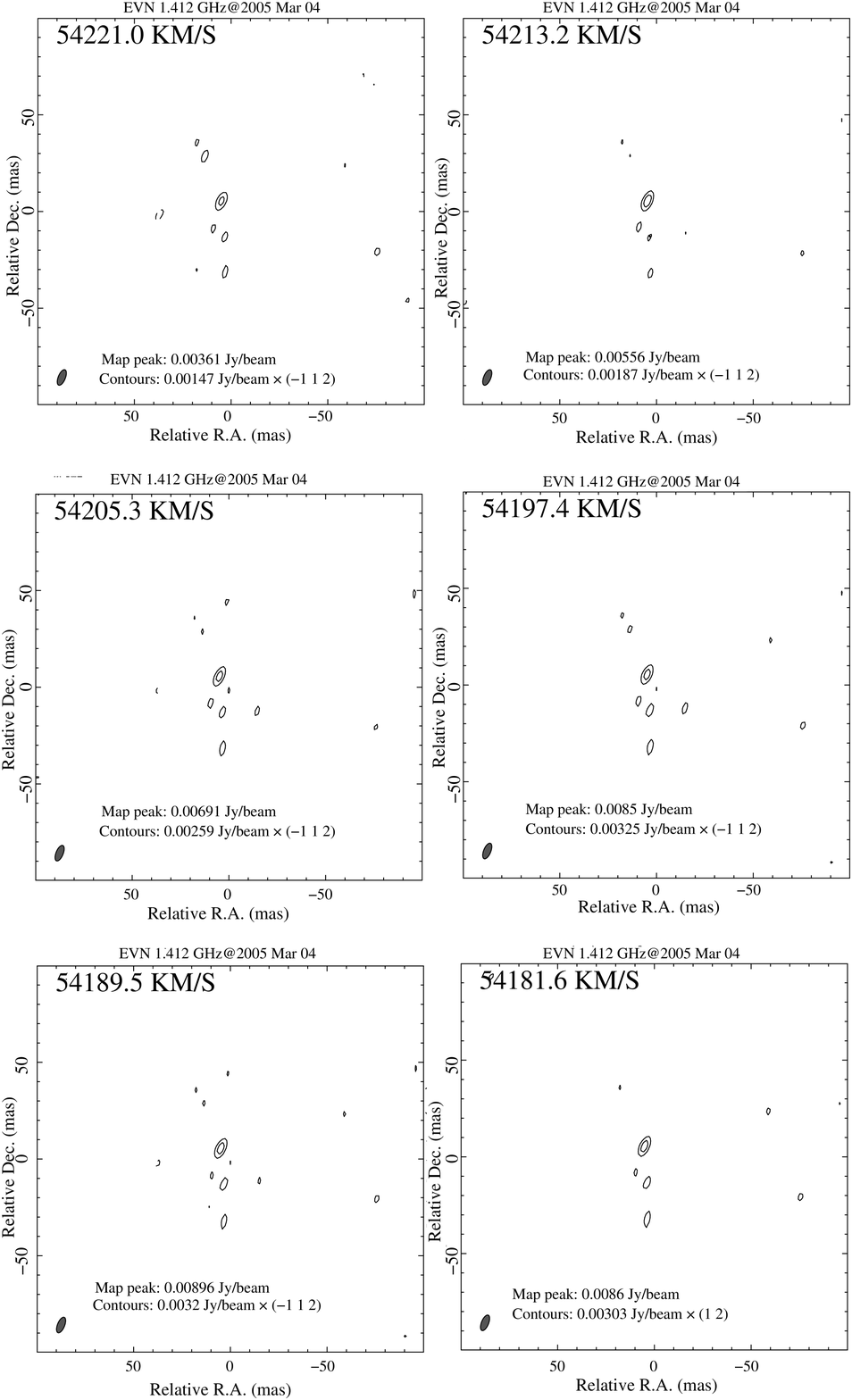}

\caption{The channel maps of IRAS 02524 observed with the EVN project GD018, the beam FWHM is 8.6 $\times$ 3.69 (mas) at $-22.3^{\circ}$ for all the maps.
      }
         \label{channelmap1}
   \end{figure*}
    \addtocounter{figure}{-1}
\begin{figure*}[htbp]
   \centering
   \includegraphics[width=17cm,height=24cm]{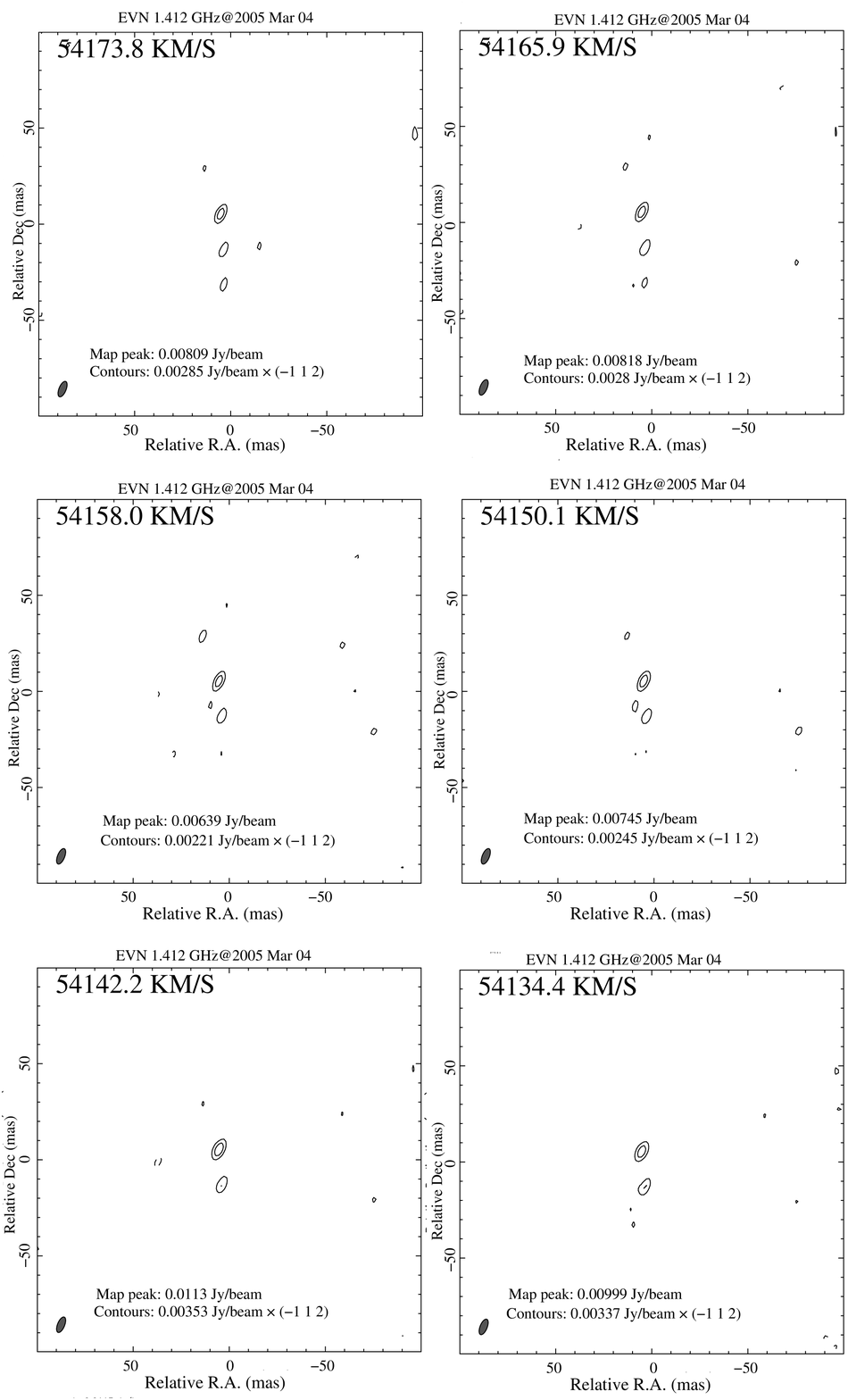}
\caption{The channel maps of IRAS 02524 observed with the EVN project GD018, the beam FWHM is 8.6 $\times$ 3.69 (mas) at $-22.3^{\circ}$ for all the maps.
      }
         \label{channelmap2}
   \end{figure*}
    \addtocounter{figure}{-1}
\begin{figure*}[htbp]
   \centering
   \includegraphics[width=17cm]{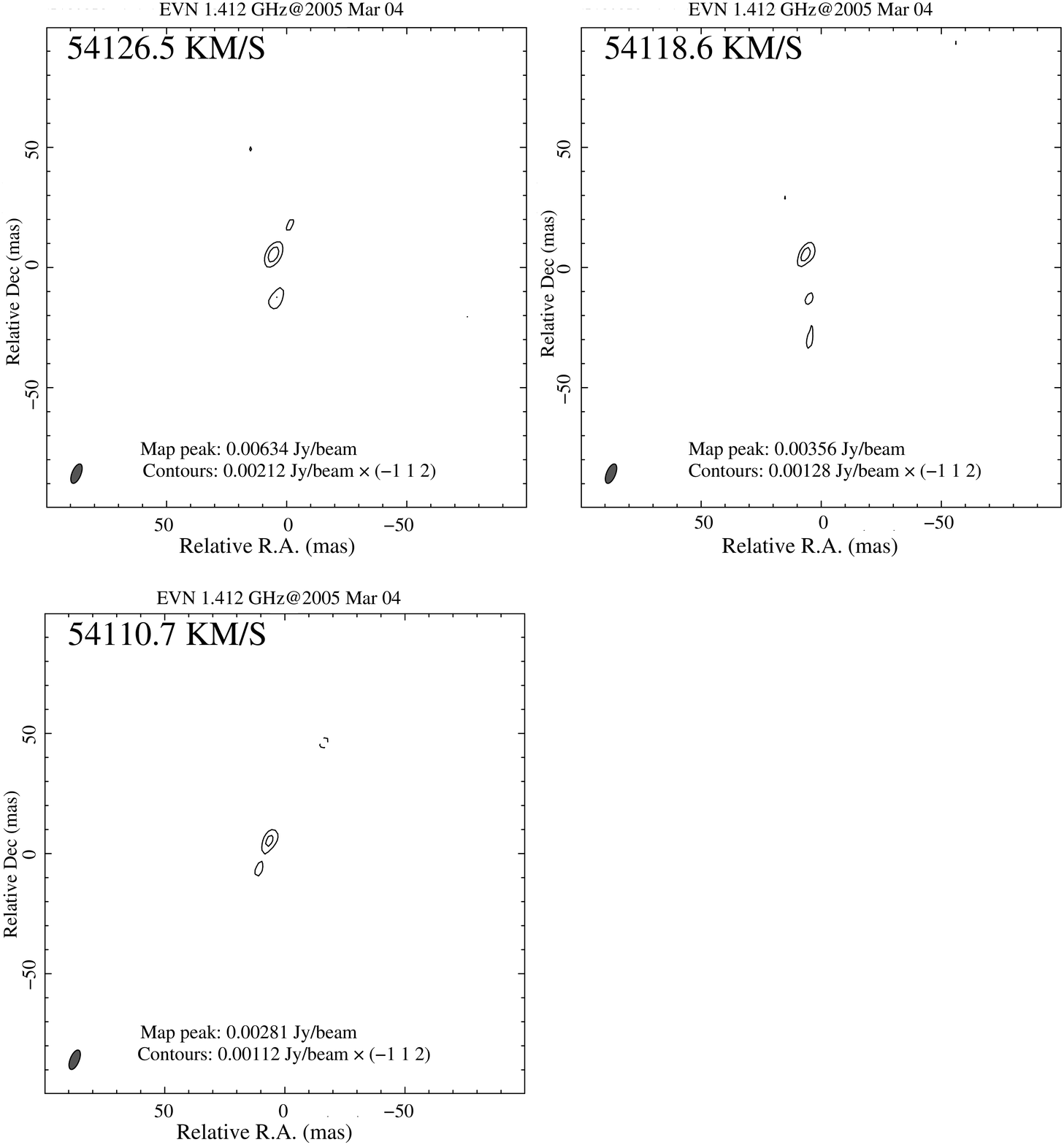}
      \caption{The channel maps of IRAS 02524 observed with the EVN project GD018, the beam FWHM is 8.6 $\times$ 3.69 (mas) at $-22.3^{\circ}$ for all the maps.
      }
         \label{channelmap3}
   \end{figure*}

\begin{figure*}[htbp]
   \centering
   \includegraphics[width=14 cm]{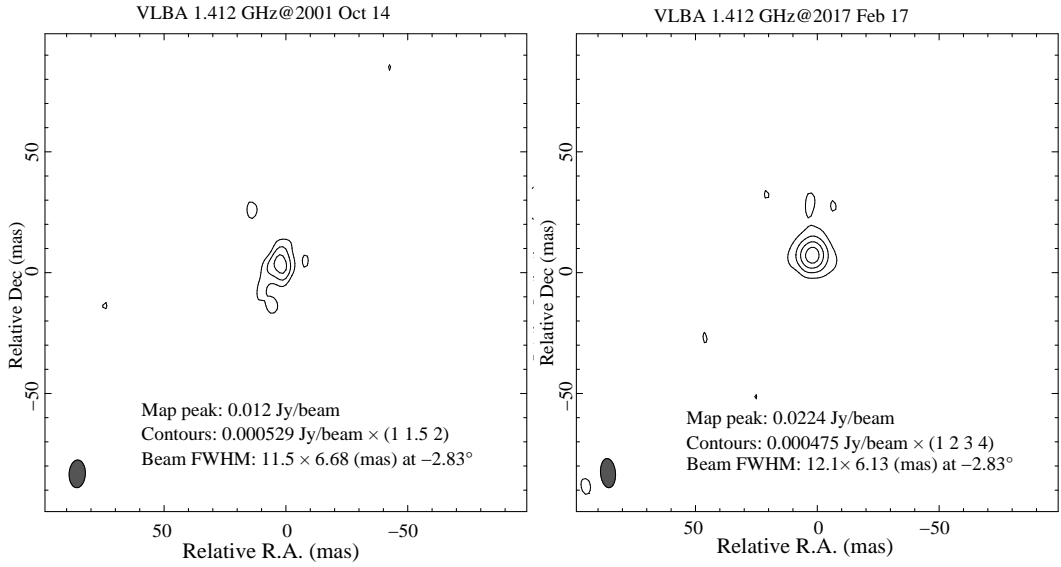}
      \caption{The OH1667 line emission of IRAS 02524 with frequency range from 1411.5 MHz to 1412.5 MHz from two VLBA observations. The frequency range corresponds to the 1667 MHz line emission from 54110.7 - 54221.0 km s$^{-1}$, the image from EVN project GD018 at this range is present in Fig. \ref{1667combin}.
      }
     \label{2vlbaline}%
    \end{figure*}

%


\begin{figure*}[htbp]
\centering
\includegraphics[width=8 cm]{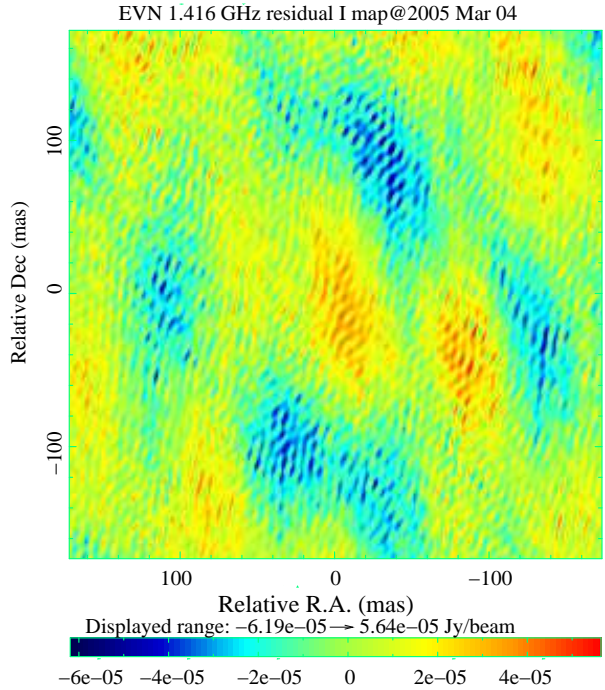}
\caption{The dirty map of the continuum emission of IRAS 02524 from EVN project GD018.
          }
\label{dirtymap}
\end{figure*}

\label{appendix}




\end{document}